# Physical probability in the Everett interpretation, and Bell inequalities[*]

*Simon Saunders*[**]

**Abstract**: I define a notion of locality $LOC$, closely modelled on Bell's principle of local causality, construed as the condition that single-case probabilities cannot be modified by actions at spacelike separation. The new principle, like Bell's, forces Bell inequalities, but with two loopholes: one is violation of measurement independence, known to Bell, but the other is non-uniqueness of remote outcomes, a loophole only for $LOC$, not for Bell's principle.

I also set out a theory of physical probability, applicable to the Everett interpretation, in which the Born rule is derived, and which therefore violates Bell inequalities. I show it is consistent with $LOC$. Surprisingly, both loopholes are exploited. I conclude not only that physical probability in the Everett interpretation involves no action at a distance, in the sense of $LOC$, but that the observed violation of Bell inequalities is powerful evidence for many worlds.

## 1. Introduction

Alice and Bob are far separated in space. There is a source of spin-systems, produced in pairs, with one from each pair sent to Alice and the other to Bob. They each measure a component of spin in one of two directions, the choice to be made freely at the time of measurement. After many repetitions of the experiment, as they work their way through many pairs of spin-systems, strange correlations are produced. Those correlations seem to defy local, causal reasoning, but they follow from quantum mechanics.

The setup is due to Bell (1964), as inspired by Einstein, Podolsky and Rosen (1935), following modifications by Bohm (1951); call it the EPRB setup. Bell argued that from local, causal principles, the probabilities for joint outcomes should satisfy certain inequalities. The experiments were actually carried out -- and here let us introduce Charlie, in the common future of both Alice and Bob, who compiles their records of outcomes; it is Charlie who concludes the inequalities are violated, with high statistical significance. Alice, Bob, and Charlie (not their real names) win the Nobel prize in physics (2022), having shown that those local, causal principles, are violated.

The challenge thus posed to Everettian quantum theory has not always been recognised. It will not do to insist that only when Charlie gathers the records are the inequalities violated, with all the measurements completed in the causal past of a single observer; we have still to explain their violation in the statistics of spacelike events. It is not enough to point to Lorentz covariance at the fundamental level, whether of the relativistic quantum state or of local quantum fields, as perhaps the issue lies with the emergent reality; and anyway, probability is not yet in the picture, when the non-locality is all about probability. The corelative step (for otherwise Lorentz covariance is spoiled) of removing collapse of the state is insufficient: if the quantum state is something physical, as in Everettian quantum theory, its collapsing by either Alice or Bob involves action at a distance, end of story; but that does not remove the apparent non-locality of probability in the EPRB setup, it removes a non-local way of explaining it. Then what does explain it? To respond, the reasoning involved presupposes a single world, and simply does not apply to Everettian theory, leaves open the question of what reasoning *does* apply, in Everettian terms, to the EPRB experiment.

[**] University of Oxford and Merton College, Oxford. Email simon.saunders@merton.ox.ac.uk

Partial answers can be found in more recent literature, for example in Timpson and Brown (2002), Bacciagaluppi (2002), Wallace (2012), Tipler (2014), and Brown and Timpson (2016). There is some consensus: like ordinary quantum mechanics, Everettian theory violates outcome Independence, one of the three key conditions used to derive Bell inequalities (this following Clauser and Horne 1974, Bell 1976); the condition is implied by what Bell called local causality, but is equivalent to another, completeness, that appears motivated rather by Reichenbach's principle of common cause – or at any rate, some doctrine on how correlations may be explained or what kinds of correlations are permissible. As argued by Brown and Timpson (2016), its failure need have no implications for action at a distance. This connects with an older tradition, according to which there is 'peaceful co-existence' with relativity, if only we give up on the demand for the explanation of correlations (Redhead 1987), as recently endorsed in Myrvold, Genovese, and Shimony (2024), which further suggests Bell's principle of local causality is a combination of Reichenbach-style explanation along with no-signalling. Everettian quantum theory is unusual, not just by virtue of many worlds, but by virtue of macroscopic entanglement – and thereby, of macroscopic non-separability -- and it is precisely for entangled states that completeness fails, even (as when Alice and Bob measure the same components of spin) Bell inequalities are satisfied. Bell's principle went wrong in failing to consider a non-separable world, for which Reichenbach's principle is inadequate.

This line of thinking appears to me true in parts, but wrong in its general direction. If correct, it suggests that many worlds is not really at issue – perhaps another kind of non-separable theory, without many worlds, may similarly violate completeness and hence Bell inequalities, with no action at a distance. I argue that on the contrary many worlds, or something close to it, is exactly what is at issue, echoing Wallace (2012 p.310). But I agree that Bell's principle of local causality went amiss -- not, however, in failing to consider the possibility of a non-separable reality, but in dispensing with any mention of cause, or dynamics.

However, the principle is easily amended, in a way that Bell might well have found congenial (I will give arguments for this claim shortly); we do not move outside of his way of thinking. Call the resulting principle $LOC$. Now for my central claim: $LOC$ can be articulated in one-world, many-world, and non-separable theories, yielding conditions (like outcome independence) in each case; but those conditions differ, and only in the one-world case, or more specifically, *only given that remote quantum experiments have unique outcomes*, are they sufficient to imply Bell inequalities. This suggests Everettian quantum theory may well be consistent with $LOC$.[1]

If we are to examine this question seriously, we will need an account of probability in Everettian theory, and specifically of objective, or as I shall call it, *physical* probability. The contrast is with subjective, or better, epistemic probability, that is bound up with knowledge, belief, and rational agency. If we are not to bring in Charlie, the account of physical probability should extend to joint probabilities for spacelike separated events. This is my other claim: it is possible to define a theory of probability, in this physical sense, available in Everettian quantum mechanics, applicable to spacelike separated events. No special premise or hypothesis is needed.[2] It is essentially a microstate-counting rule in the tradition of Boltzmann, call $\lambda$-*MANY* (the $\lambda$'s are the microstates).

This goes against the widely held view that probability in the physical, objective sense has no place in Everettian theory, as maintained by several of its advocates as well as by critics.[3] Rather than address those arguments directly, I simply construct $\lambda$-*MANY* explicitly; readers may judge for themselves. Its seeming chief weakness, that probability so defined lacks the attribute of uncertainty (as likewise ignorance, and unpredictability), may in fact be one of its strengths, for this needs to be seen in light of the distinction between physical and epistemic probability -- those missing attributes *clearly* have to do with epistemology. It may be that they come into play *only* when epistemic agents enter the picture, bringing with them features specific to epistemology, like rational choice theory and the knowing subject - and thereby, the localised subject, involving self-location and perhaps self-locating uncertainty. Insisting that physical probability must involve uncertainty in an essential way may simply be a mistake, a Baconian Idol of the Tribe.

---

[1] It is possible that relational quantum mechanics, as in Rovelli (1996) is likewise consistent with $LOC$, but it is unclear to me that it has a home for physical probability (see below).

[2] However, if we seek a derivation as opposed to an explanation of the Born rule, then a new principle is needed. See Section 3 below, and Saunders (2025).

[3] Among critics, see Saunders et al (2010 Part 4); among advocates, see Deutsch (2016), Brown and Porath (2020), Vaidman (2021). An exception is Wallace (2012), who defends the notion of physical probability, albeit there tied to epistemic probability, in such a way as to be unavailable for joint probabilities for spacelike events (Charlie must come into the picture). I have long argued that probability is physical in Everettian theory even in the absence of agents, for example in Saunders (1998, 2005, 2010).

None of which is to say that when epistemic probability *is* in play, those features do not matter. In particular, where they do arise, they had better connect in the right way, with normative effect, to physical probability. But given the remarkable Deutsch-Wallace theorem (Deutsch 1999, Wallace 2012), so long as $\lambda$-MANY yields the relevant Born rule quantities, that is assured; for that theorem shows agents rational in a requisite sense, in the face of branching, ought to conform their credence functions to the Born rule quantities for branches, insofar as they are known.

This chapter is organised as follows. Section 2 sets out the case for recasting Bell's local causality condition as $LOC$, and studies the differences that it makes to Bell theorems, both for deterministic hidden variable theories (Bell 1964) and indeterministic theories (Bell 1976) (ordinary quantum mechanics is an example of the latter). Section 3 introduces $\lambda$-MANY; it is based on Saunders (2024, 2025). Readers prepared to take it on trust may jump to the next two sections, where $\lambda$-MANY is applied to the EPRB setup and examined in terms of $LOC$, first, in Section 4, invariance under remote choices of instrument settings, yielding conditions known as parameter independence*;* second, in Section 5, invariance on remotely choosing to run the experiment, that given uniqueness of outcome forces outcome independence. Given both, and one further premise, Bell inequalities follow. But in $\lambda$-MANY, uniqueness fails; Section 5 is mainly directed to showing what the violation of outcome independence then means, unrelated to LOC.

In fact, shown at that point is that $LOC$ is respected by $\lambda$-MANY not only at the level of probabilities defined by microstate-counting, but in terms of the microstates themselves. We conclude that Everettian quantum mechanics, based on $\lambda$-MANY, is fully consistent with $LOC$. It is also consistent with non-uniqueness of remote outcomes.[4] Since it agrees with the Born rule, for certain choices of the four spin directions, it violates Bell inequalities. It gives us all that may be desired, but there is more.

[4] Not all formulations of Everett's ideas fall in this category, however, for example de Witt (1970), Sebens and Carroll (2018).

There is that further premise needed for Bell's two theorems: measurement independence, the premise that Alice and Bob's choices of measurement settings is independent of the state of their spin systems, prepared at the source (and independent of the probability distribution over those states, if they vary from one run of the experiment to the next). It forbids the seemingly conspiratorial possibility that those spin systems dictate Allice and Bob's choices of measurement settings, or that those choices change the state of their spin systems emitted at the source (or probabilities of those states), in their common past. Measurement Independence is satisfied in $\lambda$-MANY as well as in pilot-wave theory and ordinary quantum mechanics.

Given measurement independence, given uniqueness of remote outcomes, and given $LOC$, Bell inequalities follow. Since the latter are violated, we arrive at a very general argument. *Given measurement independence and $LOC$, remote quantum experiments do not have unique outcomes*. The argument is even more striking considering that $LOC$ is strictly weaker than Bell's principle of local causality.

This argument is set out in Section 6. It was already contained, although not stated explicitly, in Section 2. For failing an understanding of what it *means* for remote experiments to have non-unique outcomes, with attached probabilities, the inference on the basis of Section 2 is useless: perhaps the premises of the argument are inconsistent, perhaps probabilities of outcomes, if non-unique, make no physical sense. Sections 3-5 are needed to show that neither is true.

## 2. Local Causality and LOC

I (mostly) follow the notation and terminology of Myrvold, Genovese, and Shimony (2024). Let Alice measure the component of spin of her spin-system in one of the directions $a, a'$ (unit vectors in 3-dimensions) with outcomes $s = \pm 1$, spin-up and spin-down; and likewise let Bob measure the spin of his spin system in one of two directions $b, b'$, with outcomes $t = \pm 1$.

We suppose that any acceptable theory of the EPRB experiment will provide a complete description $\omega$ of the state of the two spin-systems as produced at the source (it is up to the theory to say what this is). Given this, and given a sufficiently detailed description of Alice and Bob's measurement apparatuses, including the parameter settings, on a given trial (that we likewise label $a, a', b, b'$ ), we suppose the theory defines joint probabilities for all pairs of outcomes $s, t = \pm 1$, for the four configurations $ab, ab', a'b, a'b'$ (so 16 real numbers bounded by 0 and 1). The question is *how does the theory do this* – what constraints among those numbers are in play, and in particular, what local, causal constraints apply? Bell proposed:

Local Causality: A theory will be said to be locally causal if the probabilities attached to values of local beables in a space-time region 1 are unaltered by *specification of* values of local beables in a space-like separated region 2, when what happens in the backward light cone of 1 is already sufficiently specified, for example by a full specification of local beables in a spacetime region 3. (Bell 1990 p.239; emphasis mine)

See Fig.1.

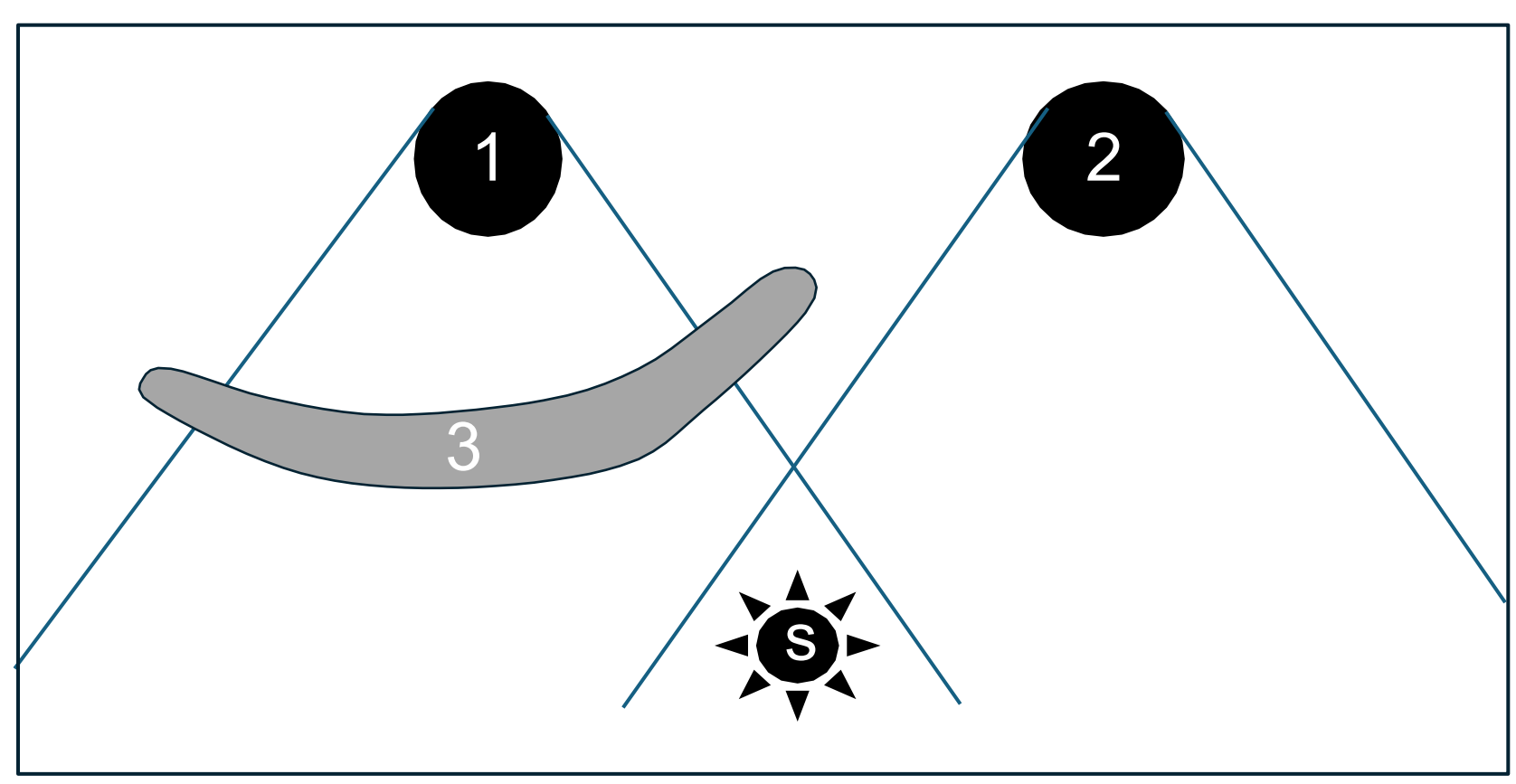

Fig 1: Alice conducts her experiment in spacetime region 1, Bob conducts his in region 2; the source of spin-systems is in region S, in the causal past of Alice and Bob. The region 3 entirely obstructs the past light-cone of 1.

By 'probabilities attached to values of local beables' Bell clearly meant single-case probabilities, defined for each trial. As bound to the causal order, we further take them to be physical probabilities (the distinction was not sharply made, in Bell's day, but he often spoke of chance, and never of credence). However, Bell's principle makes no mention of cause, so in effect it offers a definition of cause (or at least a definition of local cause). That is obviously problematic (philosophers have been trying to do that for centuries), nor is there any need for a definition of cause. All that is wanted is a criterion for when causal processes admitted by a theory (essentially, the dynamics) act locally, as regards probabilities -- we can leave it to the theory to say what those dynamical processes are, just as what the local beables are.

Bell's principle can however be easily amended, replacing 'specification of' with 'causal changes in', as italicised below, with all else the same:

> LOC: Probabilities attached to values of local beables in a space-time region 1 are unaltered by *causal changes* in values of local beables in a space-like separated region 2, when what happens in the backward light cone of 1 is already sufficiently specified, for example by a full specification of local beables in a spacetime region 3.

'Causal changes' means dynamically allowed changes, lawlike processes, according to a given theory, consistent with the full specification of local beables in region 3 (the latter also to be defined by the theory). We further take them to be changes that could, in principle, be carried out by Bob (let us grant him -- and Alice – unlimited experimental powers, as permitted by the theory in question, and consistent with the full specification of region 3). As we shall use LOC: they are changes produced by Alice and Bob, when they freely choose their instrument settings, and when they freely choose to carry out their experiments.

Bell likewise spoke of free choices of experimenters. He was criticised (in Shimony, Horne and Clauser 1976) for introducing questions about free-will into the debate. Bell (1977) defended his usage, on the grounds that 'free' meant free as opposed to dependant variables, a distinction well-founded in physics. (Myrvold, Genovese, and Shimony (2024) agrees, commending the disclaimer, but recommends the term 'exogenous' instead.) Granted, what is under laboratory control are the values of free or exogenous variables: they are typically modelled by time-dependent Hamiltonians, in control theory, exactly the right notion for $LOC$, in a Hamiltonian theory.

$LOC$ is close to local causality in another respect: invariance under a dynamical change can be expressed in terms of 'specification' of values of beables, so long as it is the *full* specification of the values of the beables involved in the dynamical change (that *just is* to specify the dynamical change). That shows local causality implies $LOC$, and is strictly stronger.

There are two more reasons to favour $LOC$. The first is that Bell frequently appealed to Einstein's dictum:

> But on one supposition we should, in my opinion, absolutely hold fast: the real factual situation of the system [S1] is independent of *what is done* with the system [S2], which is spatially separated from the former. (Einstein 1949 p.83, emphasise mine).

'What is done' denotes action, and involves dynamics, and gives support to $LOC$.

The second reason to favour LOC concerns Bell's own doubts about his definition. In his last writing on EPRB, he stated his principle of local causality again, but immediately before, he warned the reader:

> Now it is precisely in cleaning up intuitive ideas for mathematics that one is likely to throw out the baby with the bathwater. So the next step should be viewed with the utmost suspicion (Bell 1990 p.239).

No doubt he had in mind Von Neumann's impossibility proof of hidden variables – he feared the same hubris. Immediately following, he noted that his definition 'although motivated by talk of "cause" and "effect", does not in the end explicitly involve these rather vague notions' (Bell 1990 p.240). That showed his main concern was with precision rather than expunging the notion of cause. But leaving it to the theory to say what counts as a dynamical change, if the theory is any good, will provide the needed precision.

From now on we use $LOC$ rather than local causality in deriving Bell inequalities, confident that in spirit if not in letter it is the same principle. We will need to consider indeterministic as well as deterministic hidden variable theories, beginning with the former. We follow Myrvold, Ginovese and Shimony (2024), save that in the indeterministic case we make the simplifying assumption that the complete description $\omega$ of the two spin-systems is exactly the same on each trial. In that case, after a large number of trials, with various choices of spin-directions, constraints among the single-case joint probabilities will, with high probability, be found in recorded outcomes. (The requirement that $\omega$ be the same for each trial is a special case of measurement independence. Were $\omega$ to vary from one trial to the next, but be produced in accordance with a probability distribution $\rho$, measurement independence would require $\rho$ to be independent of the parameter settings. We are in effect assuming $\rho$ is a delta-function, non-zero only for $\omega$.)

Under these assumptions, we assume a theory of the EPRB experiment will define mappings from pairs of parameter settings and pairs of outcomes to real numbers (probabilities) for a single trial. There are 16 such inputs so 16 such numbers. We write $p_{a,b}(s,t|\ \omega\ )$ as the joint probability for the pair of outcomes $s,t$ when Alice's measurement settings are set to $a$ and Bob's to $b$ (4 joint probabilities, for 4 pairs of outcomes), with similar expressions for the three other pairs of instrument settings. (Although only one pair of instrument settings is actually chosen, for each trial, just insofar as the choice is exogenous, the theory must specify the probabilities that would obtain, were other instrument settings chosen.)

We assume elementary probability theory. The marginal probabilities for Alice and Bob's measurements are

$$p_{a,b}(s|\ \omega\ ) = \sum\nolimits_t p_{a,b}(s,t|\ \omega\ )$$

$$p_{a,b}(t|\ \omega\ ) = \sum\nolimits_s p_{a,b}(s,t|\ \omega\ )$$

where

$$\sum\nolimits_s p_{a,b}(s|\omega) = \sum\nolimits_t p_{a,b}(t|\omega) = 1$$

and similarly for other pairs of measurement settings.

From $LOC$, and supposing Bob is free to choose his instrument setting, whether $b$ or $b'$, consistent with the full specification of local beables in region 3, this action must leave Alice's probabilities unchanged. This must be so for each of her instrument settings, therefore:

$$p_{a,b}(s|\omega) = p_{a,b'}(s|\omega) \qquad (1)$$

$$p_{a',b}(s|\omega) = p_{a',b'}(s|\omega).$$

Similar identities apply to Bob's probabilities. This set of conditions is parameter independence. Were one of them systematically violated, on repeated measurements, since Bob directly controls his instrument settings and we suppose $\omega$ is always the same, probabilistic superluminal signalling would be possible.

We distinguish Bob's action in setting his instrument to either $b$ or $b'$, from his action in actually carrying out his measurement. The latter too is a free action – he may or may not be so inclined -- so $LOC$ applies to it as well: if Bob chooses to perform his experiment, his action should leave Alice's probabilities unchanged. *Given that the outcome of his experiment is unique,* say $t = +1$ (and leaves all data for region 3 unchanged), it must count as a causal, dynamically allowed change, in any theory guaranteeing uniqueness. Therefore, under $LOC$, Alice's probabilities should be unchanged. Assuming the unique outcome $t = +1$ has non-zero probability, this may be written in terms of the conditional:

$$p_{a,b}(\ s|\omega) = p_{a,b}(s|t = +1, \omega). \qquad (2)$$

Since by assumption the probabilities for $t = \pm 1$ sum to one, from the definition of conditional probability, if (2) holds a similar identity must hold for $t = -1$, if of non-zero probability. The same should follow given Alice's measurement setting $a'$, as also on interchanging the roles of Alice and Bob to define identities among Bob's probabilities. This set of conditions is outcome independence.

Given uniqueness of remote outcomes, outcome Independence follows directly from $LOC$. It has nothing to do with the explanation of correlations; it is an expression of no distant action on single-case probabilities, the same as with parameter independence. The difference is that even if identities like (2) are systematically violated, so long as the unique outcome is not controllable by Bob, no superluminal signalling to Alice is possible.

All of the foregoing applies if we use Bell's local causality in place of $LOC$, save in one respect: in the derivation of outcome independence there is no longer a proviso about uniqueness of remote outcomes. For even if Bob remotely produces, somehow, both outcomes, *it is always possible to specify just one of them* – whereupon, if Alice's probabilities are to be invariant under this specification, we obtain Eq.(2), and hence outcome independence, regardless. The uniqueness or otherwise of Bob's outcome is irrelevant. Everettian quantum theory therefore violates local causality (this point is agreed – see, for example, Brown and Timpson 2016), but in light of LOC, it seems it fails on a technicality, and an unintended one at that.

From our dynamical perspective, it is far from a technicality: it amounts to the requirement of invariance under a partial specification of Bob's measurement process, as a process yielding a unique outcome, as in ordinary quantum mechanics. But that is just collapse of the state! To demand invariance under what in Everettian theory is a violation of physical law (collapse) is wrong-headed, even if confined to a remote region of space. In light of Bell's comment on Everett (that we shall come on to shortly), it is hard to believe it was intended.

For future reference, we need the symmetric conditions:

$$p_{a,b}(s,t|\omega) = p_{a,b}(s|\omega)p_{a,b}(t|\omega). \qquad (3)$$

For values of $s, t$ for which (3) is non-zero, (3) implies (2), and vice versa. Similar conditions follow for other choices of parameter settings. This system of identities is completeness.[5] Although it is equivalent to outcome independence, its connection with $LOC$, or indeed with local causality, is less clear. It has been linked to Reichenbach's common-cause principle, with $\omega$ the common cause in the past of the events $s, t$, for (3) conditionalizes on $\omega$ alone, produced in the past; so it is a kind of causal constraint, in terms of the past. In contrast the identities in outcome independence, like Eq.(2), conditionalize on spacelike events, and concern distant actions.

[5] Terminology for conditions (2), (3) varies wildly. Eq.(3) first appeared in Clauser and Horne (1974) but was there unnamed; 'completeness' was introduced in Jarrett (1984). Myrvold et al (2024) calls (3) and allied conditions 'outcome independence', citing Shimony (1986).

Given parameter independence, outcome independence, and measurement independence, Bell inequalities, and specifically CHSH inequalities, follow immutably. The relevant conditions can all be translated into ordinary quantum mechanics quite easily, replacing $\omega$ by the quantum state, introducing a tensor product structure for the two regions 1 and 2, and computing probabilities using the Born rule. Alice and Bob are free to choose local unitary transformations (they control local, time-dependent Hamiltonians, 'free parameters' according to Bell, 'exogenous variables' according to Myrvold et al). Measurement independence is then assured (we may assume the quantum state is reliably the same, as produced by the state-preparation device on each trial). Parameter independence follows easily -- this is the quantum no-signalling theorem, that first came to light in Bell's work. But outcome independence is violated when the quantum state of the two spin systems is an entangled state of spin. We are talking about ordinary quantum mechanics; remote quantum experiments have unique outcomes, so this is action at a distance, a change in Alice's probabilities that has resulted from a remote, lawlike change, as produced by Bob's free choice to carry out his experiment. Ordinary quantum mechanics thus violates $LOC$.

This is exactly as expected. But now, also expected is that Everettian quantum theory can exploit the loophole in this argument, and still satisfy $LOC$, even if it violates outcome independence, namely: if remote experiments have non-unique outcomes. Many worlds may thus well make a difference. Bell himself suspected as much, despite his unfavourable view of Everett's ideas, writing shortly before his untimely demise:

> The 'many world interpretation' seems to me an extravagant, and above all an extravagantly vague, hypothesis. I could almost dismiss it as silly. And yet…it may have something distinctive to say in connection with the 'Einstein Podolsky Rosen puzzle', and it would be worthwhile, I think, to formulate some precise version of it to see if this is really so. (Bell 1986 p.194.)

We are attempting to meet his challenge, first, in the next section, making precise the notion of probability, and second, in the two subsequent, applying it to the EPRB experiment.

Before that, we review Bell's other, earlier (1964) theorem, for a deterministic hidden variable theory. We will need this in the sequel as well. To distinguish this kind of theory from the previous (indeterministic) case, we denote the complete state of the spin systems by $\lambda \in \Lambda$, rather than $\omega$, where $\Lambda$ is a suitable measure space equipped with a probability distribution $\rho(\lambda)$. Otherwise, we use the same notation as before, now assuming probabilities conditional on $\lambda$ are always 0s and 1s:

$$p_{a,b}(s,t|\lambda) \in \{0,1\};\ \lambda \in \Lambda \tag{4}$$

and similarly, for $a, b'$, etc., for all values of $s$ and $t$ and every $\lambda$.

Nontrivial probabilities for outcomes are given by averaging over the $\lambda$'s with respect to $\rho$, on the assumption that many measurements are made. In this way we arrive at probabilities:

$$p_{a,b}(s,t|\rho) = \int_{\Lambda} p_{a,b}(s,t|\lambda)\rho(\lambda)d\lambda. \tag{5}$$

These are the analogues of the probabilities taken as basic in our previous treatment, conditional on $\omega$. The difference is that before they were single-case probabilities, defined at the time of measurements, attached to outcomes in region 1 and 2 of Fig.1; whereas now, conditional on $\lambda$ at the time of measurement, the single-case probabilities are all 0s and 1s. Correspondingly $LOC$ applies to those 0s and 1s, rather than the non-trivial probabilities as given by (5). (A joint probability of the form Eq.(5) is more straightforwardly interpreted as an average of 0s and 1s over repeated measurements, rather than as a single-case probability.)

Such a theory is *factorizable* if

$$p_{a,b}(s,t|\lambda) = p_a(s|\lambda)p_b(t|\lambda) \tag{6}$$

and likewise for other parameter settings, for all pairs of outcomes and for every $\lambda$. From this and (5), and assuming measurement independence, Bell inequalities follow.

Measurement independence now applies to the $\lambda$'s as well as $\rho, \Lambda$; the hidden variable $\lambda$ must be independent of the choice of instrument settings, neither dictating them (superdeterminism) nor dictated by them (retrocausation).[6] Equally the measure $\rho, \Lambda$ in (5) must be independent of the measurement settings.

Factorizability follows from parameter independence and outcome independence, both conditional on $\lambda$. But now the second of these is automatic, as this is a deterministic one-world hidden variable theory. If $\lambda$ determines Bob's unique outcome, Alice's probabilities (1s and 0s) for her outcomes, conditionalizing on $\lambda$, must be the same as conditionalizing on $\lambda$ and that outcome.

Parameter independence, conditional on $\lambda$, is another matter. But this clearly follows from $LOC$:

$$p_{a,b}(s|\lambda) = p_{a,b'}(s|\lambda) \tag{7}$$

---

[6] Nor constrained in other ways: see Palmer (2024) for a third option. For a study of retrocausation in this context, see Price and Wharton (2015).

(and similar, interchanging Alice and Bob). Factorizability now follows in turn. Given measurement independence, so do Bell inequalities. Bell (1964) called (7) 'the vital assumption', with a footnote quoting Einstein (the quotation already given). It clearly follows from local causality as well.

Any deterministic hidden-variable theory that reproduces quantum probabilities and satisfies measurement independence must violate identities like (7), for certain choices of spin directions, and hence violate both local causality and $LOC$. Violation of parameter independence conditional on $\rho$ would be fatal, as from measurement independence $\rho$ is always the same, so superluminal signalling would be possible; but violating parameter independence conditional on $\lambda$ has no such implication, for now $\lambda$ is uncontrollable, and the impact of Bob's change in instrument settings on Alice's probabilities depends on $\lambda$. Averaging over $\lambda$ on repeated measurements masks the action at a distance for each measurement, with the result that parameter independence conditional on $\rho$ yet holds. It is a somewhat torturous circle of ideas.

Yet it answered the question Bell had previously posed in his (1966): he knew that quantum probabilities could be reproduced in a deterministic hidden-variable theory, namely in pilot-wave theory (he cited Bohm's 1952 papers), but that explicitly involves action at a distance; might there be some other theory of this type, that is free of it? His answer, in Bell (1964),[7] was negative; any such must violate (7), just like pilot-wave theory.

Note, in conclusion, the curious dualism between the two Bell theorems (for indeterministic theories like ordinary quantum mechanics, and for deterministic theories like Bohmian mechanics): in each individual trial, outcome independence is violated (satisfied) and parameter independence is satisfied (violated) in indeterministic (deterministic) hidden variable theories, like ordinary quantum mechanics (Bohmian mechanics). Wiseman (2014) comments on this. We shall come back to it in Section 4.

## 3. The theory $\lambda$-MANY [8]

We assume the basic framework of Everettian quantum mechanics: there exists a unique vector state $\psi \in \mathcal{H}$, the universal wave-function, the superposition principle has unlimited scope, and the dynamics is always unitary. No more detail is needed. Here $\psi$ may have any (real) amplitude and (complex) phase.

[7] It seems there was no exciting retrocausation: Bell's earlier submission was lost by the editorial office.

[8] Here I closely follow Saunders (2025); the main difference from my (2024) is that there is no longer the constraint that microstates decohere.

We consider expansions of $\psi$ into *microstates*: orthogonal vector states of equal Hilbert-space norm (so equal real amplitude). For any such expansion, denote by $\Lambda_\psi^n$ the set of microstates, for some integer $n$ ($n$ must be finite since $\psi$ must have finite norm). We define probabilities for projectors $P$ on $\mathcal{H}$ in a series of steps.

To begin, suppose there is an equiamplitude expansion of $\psi$ in microstates that are all eigenstates of $P$. That is, suppose:

$$\psi = \xi_1 + \ldots + \xi_n \ ; \quad \|\xi_j\| = \|\xi_k\|; \quad P\xi_j = \xi_j \text{ or } 0; \quad j,k = 1, \ldots, n.$$

In such a case, we shall say $\Lambda_\psi^n$ is *adapted* to $P$. Suppose $m$ microstates lie in the range of $P$, so $n - m$ lie in $I - P$; we then say that the probability of $P$ is $m/n$, because its frequency in $\Lambda_\psi^n$ is $m/n$. It trivially follows that $\Lambda_\psi^n$ is adapted to each of the projectors $P_{\xi_k}, k = 1, .., n$, and that each has probability $1/n$. Implicit, then, in this picture of probability, is the condition that microstates be equiprobable.

We can rephrase the rule in terms of states. Let the superposition of $m$ microstates equal $P\psi$, and of $n - m$ equal $(I - P)\psi$: if the microstates have equal probability, then the probability of $P\psi$ is $m/n$. Similarly, the probability of $P_{\xi_k}\psi = \xi_k$ is $1/n$. (In this form, it applies to projectors of a slightly more general nature – when, for example, individual microstates are not eigenstates, but there is a partitioning of $\Lambda_\psi^n$, with the sum of vectors in each partition an eigenstate of $P$; but this need not concern us here.)

The Born rule for $P$ in the state $\psi$ agrees with our microstate-counting rule:

$$\frac{\|P\psi\|^2}{\|\psi\|^2} = \frac{\|P(\xi_1 + \ldots + \xi_n)\|^2}{\|\xi_1 + \ldots + \xi_n\|^2} = \frac{\|(\xi_1 + \ldots + \xi_m)\|^2}{\|\xi_1 + \ldots + \xi_n\|^2} = \frac{m}{n}.$$

The same equation follows for any other equiamplitude expansion $\Lambda_\psi^{n'}$ of $\psi$ into $n'$ microstates adapted to $P$, yielding probability $m'/n'$. Therefore $m'/n' = m/n$.

Call this theory $\lambda$-MANY (explanation: the $\lambda$'s are the microstates). For projectors like these, the theory is simple. But what of expansions that are not adapted to $P$? They yet place *bounds* on the probability of $P$ – an upper-bound, defined by the number of microstates that have eigenvalue 0, and a lower-bound, defined by the number of microstates that have eigenvalue 1. There is a grey zone in between, consisting of microstates that are not eigenstates of $P$ - call them Schrödinger-cat states for $P$. The larger this interval, the less informative the expansion, but since all the microstates are equiprobable, these bounds on probabilities can never contradict one another. This can be proved directly: for any expansion, the interval between upper and lower bound for the probability of $P$ must always contain the Born-rule quantity, so they can never be disjoint.[9] Call such pairs of bounds *imprecise probabilities*.[10]

This picture of probability is clearly akin to frequentism, indeed finite frequentism, in (classical, one-world) philosophy of probability, where likewise the fraction of an appropriate finite ensemble, that has property $P$, is the probability of $P$ in that ensemble. Likewise, the members of the ensemble are implicitly or explicitly taken as equiprobable. Likewise, there may be borderline cases, so imprecise probabilities may enter as well. The difference lies in the nature of the ensemble: classically, it is spread out across space and time, involving many repeated measurements; quantumly, it is spread out across a superposition at a single place and time, involving a single measurement.

To make the picture vivid, consider the balls-in-urn (with replacement) model of probability, beloved of frequentists. Let the balls be marked either blue or green. Classically, the fraction of balls in the urn that are green gives the probability that a ball drawn randomly from the urn is green. Quantumly, all the balls are drawn from the urn, and measuring their colour partitions them into those that are green, and those that are blue, with probability of green given by the fraction that are green, and similarly for blue. The partitioning is done by a quantum measurement apparatus under the unitary dynamics. One or more of the balls may be ambiguous as to colour – we have a Schrödinger cat. There may be no measurement possible whose outcomes correspond to individual balls (the differences between the balls being too tiny to discriminate) -- microstates are not Everettian worlds, or branches, which (for the sake of argument) we suppose must differ macroscopically. Worlds may have different probabilities, because the numbers of microstates that superpose to give those worlds differ, and because the microstates are equally probable.[11] Epistemic probability comes into the picture if we imagine there are observers, witnessing the draw, finding themselves partitioned as well, some finding themselves in the green partition, some in the blue. At least momentarily, they will be uncertain as to which partition they are in, whether the green or the blue (they will have 'Vaidman uncertainty', Vaidman (1998)).

More is needed to obtain probabilities as real and not just rational numbers. To that end, suppose $\dim \mathcal{H} = \infty$, and consider projectors $P$ on $\mathcal{H}$ with $\dim P = \dim(I - P) = \infty$. They are otherwise arbitrary, defined without any reference to $\psi$. Projectors like these are needed to describe any system with a continuous degree of freedom, like the position of an atom in space; a projector $P$ onto the closure $\Delta$ of any open set in space, no matter how small, has infinite dimension (the Hilbert space $L^2(\Delta, dx)$ is infinite-dimensional), and likewise $I - P$. Arguably, *any* realistic physical system falls into this category, as involving at least one continuous degree of freedom.[12] (This need for projectors of infinite dimension is often concealed. If we separate out spin-degrees of freedom in terms of a tensor product structure, we have projectors of the form $P \otimes I$, where $P$ is of finite and perhaps very small dimension; if we ignore the centre-of-mass degrees of freedom, spin-systems can be treated entirely in the framework of finite-dimensional Hilbert space. But spin-systems have spatial properties -- or at least, the ones in the EPRB setup do -- so the underlying projectors are really of the form $P \otimes I$, as likewise their complements $(I - P) \otimes I$, which are always infinite-dimensional.)

---

[9] The proof is easy; see Saunders (2024)

[10] Following e.g. Bradley (2019). This literature is almost exclusively concerned with epistemic probability (and by default, restricted to a single world).

[11] The suggestion that branches be considered equiprobable, made in Graham (1973), is of course disastrous for Everettian quantum theory, but fortunately has no good argument. See Saunders (2024) for a more plausible variant on Graham's branch-counting rule; this also fails, refuting Khawaja (2023).

I state without proof two key properties of equiamplitude expansions in Hilbert space:

EE1 If $\dim \mathcal{H} \geq 3$, for any vector $\psi \in \mathcal{H}$ and for any $n$, $2 \leq n \leq \dim \mathcal{H}$, there exists a continuous infinity of equiamplitude expansions of $\psi$ in $n$ microstates

EE2 If $\dim \mathcal{H} = \dim P = \dim(I - P) = \infty$, then for any state $\psi \in \mathcal{H}$ and any $n \geq 2$, there exists an equiamplitude expansion of $\psi$ into $n$ microstates, at most one of which is not an eigenstate of $P$.

The first can be proved by simple induction (Saunders 2025), the second by explicit construction. We shall still say finite expansions of the form EE2 for large $n$ are adapted to $P$, ignoring the Schrödinger-cat state, as $n$ can be taken as arbitrarily large, and in the limit $n \to \infty$, the probability thus defined is precise. In this way we obtain a vastly expanded class of projectors for which precise probabilities can be defined – as real numbers.

This use of the infinite limit is common to all forms of frequentism that aspire to define probabilities as real and not just rational numbers, but there are two important differences from the classical case. Classically, $n$ is the number of trials of an experiment; of course, the $n \to \infty$ limit is unphysical, but even moderately large numbers are unphysical – a billion tosses of a single coin is unphysical. The $n \to N_A$ limit, where $N_A$ is Avogardro's number, is definitely unphysical, for any kind of classical trial. But in the quantum case it poses no difficulty; nor does the limit

$$n \to {N_A}^{({N_A}^{N_A})}$$

and so on, without bound, given that $\dim \mathcal{H} = \infty$. The superposition that results, no matter of how many microstates, is always a representation of the same physical state $\psi$ at the same instant of time, for the same single experiment.

[12] This point is open to challenge, particularly in quantum gravity; but it seems appropriate that the physics of the continuum, for physical probability as for space, should stand or fall together. Still, $\lambda$-MANY applies, but 'most' projectors would have only imprecise probabilities (for a Hilbert space with dimension of order the Beckenstein bound, evidently the imprecision will be vanishingly small).

The second and even more important difference is that classically, the frequency of $P$ in an infinite ensemble, as defined by taking the limit $n \to \infty$,[13] may differ arbitrarily from its frequency in any finite sub-ensemble. If the subensemble is chosen, somehow, at random, the frequency of $P$ in the sub-ensemble will *probably* equal its limit frequency -- this the law of large numbers -- but probability will then depend on this notion of randomness. In the quantum case, none of this applies. The limiting frequency of $P$ necessarily satisfies the bounds placed on it by any expansion in microstates $\Lambda_{\psi}^{n}$ , for any $n$ – bounds which, by EE2, for expansions adapted to $P$, have imprecision of at most $1/n$. Those bounds are not *probably* correct, they are correct simpliciter. And no *random* selection of sub-ensemble is involved, in the choice of an equiamplitude expansion; on the contrary, it is fine-tuned to $P$.

Consider now the scope of $\lambda$-MANY. The limit of the sequence of imprecise probabilities for $P$, as secured by EE2, is a real number, equal to the Born rule quantity for $P$ in the state $\psi$. This includes the Born rule in ordinary quantum mechanics as applied to any macroscopic experiment; since apparatuses have spatial degrees of freedom, projectors representing their properties must satisfy EE2. Therefore $\lambda$-MANY includes the Born rule as ordinarily applied. It further extends to projective probabilities as defined in decoherence theory – projectors 'diagonal in the decoherence basis' (expansions of $\psi$ adapted to such projectors will likewise be diagonal in this basis) – and in this way includes more sophisticated uses of the Born rule, including decoherence-based Everettian quantum theory.[14] But $\lambda$-MANY goes well beyond this too, for it applies equally to isolated microscopic systems, to any quantum system with at least one continuous degree of freedom, whether or not there is any decoherence in play. However, the probabilities thus defined are always instantaneous probabilities. In questions of how they align over time, and what kinds of projectors can be so aligned, and whether probabilities can be assigned to sequences of projectors as histories, that are assigned probabilities in turn – all this brings in the dynamics and ultimately decoherence theory.

---

[13] Called 'hypothetical frequentism', in the terminology of Hájek (1996) – but as we have just seen, classically, even ensembles that are only moderately large are unphysical.

[14] There is obviously more to say about POV measures, and expectation values of self-adjoint operators more generally. To be continued.

All of this is promissory, however, for integral to $\lambda$-MANY is the equality of probabilities of states of equal Hilbert space norm. However natural, it is in effect an *assumption*.

Can we not appeal to the Deutsch-Wallace decision-theory approach? Hardly, or not with the intended scope. In any case, we are attempting to hold physical and epistemic notions of probability strictly apart. Might we simply posit the Born rule, to hold with complete generality? But that goes well beyond $\lambda$-MANY, and renders it pointless. Might we posit the Born rule for decohering branches? But that is not enough, for microstates are not in general branches, and may not decohere; and again, it makes $\lambda$-MANY redundant for the EPRB experiment, which of course involves decohering branches.

It seems we are back to assuming equality in norm implies equality in physical probability. However, we can do better. From a mathematical point of view, a probability theory is defined by a finite measure on a space of elements, where the latter has the structure of a Boolean algebra. In set - theoretic terms, this algebra is defined by the complement, intersection, and union of sets; in logical, predicative terms, it is defined by negation, conjunction, and disjunction. Probabilities are additive only for *disjoint* elements X, Y (sets of null intersection, predicates which cannot jointly be satisfied); disjointness is essential to probability. We are interested in single-case physical probabilities, which in a number of respects are hypothetical entities, whose existence has been widely doubted: in the circumstances it may be appropriate to subject them to a new hypothesis. I propose:

INVARIANCE The physical probability of X cannot be altered by a causal change in Y, disjoint from X.

On translating into a physical theory, the hypothesis is clearly defeasible. As before, 'causal change' means allowed dynamical change, in accordance with physical laws, as defined by a given theory. The principle is an abstraction from $LOC$, replacing the relation 'spacelike' by 'disjoint', to hold throughout the causal change.

Translated into Everettian quantum mechanics, the Boolean algebra is defined by some (any) expansion of the universal state into orthogonal vector states. Disjointness is given by orthogonality (for the details, see Saunders 2025), and X, Y etc are vectors in the expansion, and superpositions of such.

Suppose now that all the components in the expansion are assigned probabilities, summing to unity, according to some definite rule. Suppose (in keeping with Everett) that causal, dynamically allowed changes are all unitary transformations. Then INVARIANCE implies vectors of equal norm have equal physical probability.

To sketch the proof, let $\varphi$ occur in some orthogonal expansion $\Sigma$ of $\psi$, with physical probability $\mu_{\psi_\Sigma}[\varphi]$; INVARIANCE requires that if $\varphi$ is unchanged by a unitary transformation $U$, its probability is unchanged:

$$\varphi \in \Sigma;\ U\varphi = \ \varphi \Rightarrow \mu_{U\psi_{\mathrm{U}\Sigma}}[\varphi] = \mu_{\psi_\Sigma}[\varphi]\ .$$

Using this and additivity of probabilities, it is possible to construct a sequence of unitaries whose net effect is to interchange the amplitudes and probabilities assigned to two orthogonal vectors, in a superposition, with all other orthogonal vectors unchanged. If the amplitudes of those two vectors are the same, the total state is the same, so their probabilities must be the same (see Short 2023, Saunders 2025):

$$\varphi, \eta \in \Sigma\,, \qquad \|\varphi\| = \|\eta\| \ \Rightarrow \mu_{\psi_\Sigma}[\varphi] = \mu_{\psi_\Sigma}[\eta].$$

It follows, when $\Sigma$ is an equiamplitude expansion of $\psi$, that the states in the expansion (microstates) all have equal physical probability – the needed justification for $\lambda$-MANY.

Given that $\lambda$-MANY satisfies INVARIANCE quite generally, with X given by $P\psi$ for any $P$ (this needs further work; see Saunders 2025), we have a proof of the Born rule in our extended sense. Conversely, it is easy to see that the Born rule satisfies INVARIANCE.

## 4. $\boldsymbol{\lambda}$-MANY and Parameter Independence

Now consider $\lambda$-MANY as applied to the EPRB experiment. We anticipate that it violates only outcome independence, as does ordinary quantum mechanics, but since remote experiments do not have unique outcomes, this is no longer an expression of action at a distance. Parameter independence is surely satisfied. In fact, we find that parameter independence presents interesting questions in its own right, the subject of this section. Outcome independence is considered in Section 5.

$\lambda$-MANY is applied to EPRB in the same way as ordinary quantum mechanics, with $\omega$ given by the quantum state $\psi$, but replacing the measurement postulates by microstate counting, with no collapse of the state. The condition that Alice and Bob are causally isolated is made out as suggested before, by use of a tensor-product $\mathcal{H} = \mathcal{H}^A \otimes \mathcal{H}^B$, with Hilbert space $\mathcal{H}^A$ for Alice and her apparatus and spin system, $\mathcal{H}^B$ for Bob and his apparatus and spin system, and by the condition that Alice and Bob freely control their local unitaries, of the form $U\otimes I$ for Alice, and $I\otimes U$ for Bob.

Measurement outcomes (certain configurations of their apparatuses) are represented by projectors, whereby $P_s^a$ represents Alice's measurement of spin in direction $a$ with outcome $s$, and similarly $P_t^b$ for Bob's measurement in direction $b$ with outcome $t$. Thus $P_s^a\otimes P_t^b$ represents Alice's measurement outcome $s$ and Bob's measurement outcome $t$ on measuring spin in directions $a, b$ respectively. Since representing macroscopic pointer positions, if not entire configurations of apparatuses, these projectors satisfy EE2.

Given this, from the vantage point of $\lambda$-MANY, it is hard to see the fraction of microstates in $P_s^a\otimes P_t^b$ in state $\psi$, the joint probability for the two events, as any the less well-defined than the fraction in the projector $P_s^a \otimes I$, or $I\otimes P_t^b$. All three are projectors, their probabilities as fractions equally meaningful, no matter that one of them is a joint probability for spacelike events.[15]

Parameter independence is obviously satisfied. Let Alice measure spin in direction $a$, and consider outcome s, represented by the projector $P_s^a$ on $\mathcal{H}^A$, and by the projector $P_s^a\otimes I$ on $\mathcal{H}$, both of infinite dimension. Any equiamplitude expansion adapted to $P_s^a\otimes I$ will do to determine its probability – and is clearly independent of Bob's instrument setting. A fortiori, no change in Bob's settings by a local unitary can change this expansion, or the fraction of microstates in $P_s^a\otimes I$.

The point is equally obvious from the Born rule. However, caution is needed, *since it is obvious in pilot-wave theory as well.* Pilot wave theory satisfies parameter independence conditional on $\rho$, as in Eq.(5), but not conditional on $\lambda$, Eq.(7) -- the latter showing the underlying action at a distance in the theory. It is masked by averaging over the $\lambda's$ using $\rho$ in Eq.(5), what I earlier called a torturous circle of ideas.[16]

But now it seems the same must be true of the individual microstates in $\lambda$-MANY, in expansions adapted to Alice and Bob's experiments. For those microstates *also* assign measurement outcomes deterministically, so satisfy outcome independence, conditional on the individual microstate; if they satisfied parameter independence too, in the sense of Eq.(7), factorizability would follow, so that on averaging, Bell inequalities would be forced. It seems individual microstates in $\lambda$-MANY must fail parameter independence, just as do individual $\lambda$'s in pilot-wave theory.

[15] It is otherwise if all probabilities are ultimately epistemic. Brown and Timpson (2016 p.112) suggests as much 'it is only in [Charlie's] region that measurement outcomes for the individual measurements on either side can become definite with respect to one another'; prior to that the Born rule yields 'only formal statements, regarding what one would expect to see, were one to compare the results of measurements on the two sides'.

[16] Greatly ameliorated if developed as a theory in which quantum non-equilibrium is possible, as in Valentini (2025). However, non-equilibrium is proving fiendishly hard to find.

A similar question arises in questions of ontology in Everettian quantum theory; worlds are nonlocal in some sense, but the superposition of worlds is local; Everett's relative states are nonlocal in some sense, but their superposition is local. Does the same apply to microstates in $\lambda$-MANY, and if so, is the kind of nonlocality at issue action at a distance?

To answer this question, let us define a new one-world hidden variable theory, call $\lambda$-ONE. We suppose we have an expansion of $\psi$ adapted to $P$ (for the EPRB set-up, this is of the form $P_s^a \otimes P_t^b$), and thus a set of microstates $\Lambda_\psi^n$ for some large $n$. The new theory then says that preparing the state $\psi$, in fact just one of the microstates $\lambda \in \Lambda_\psi^n$ is produced at the source, chosen at random (meaning with uniform probability distribution $\rho = 1/n$ on $\Lambda_\psi^n$). This microstate assigns $P$ either 0 or 1 (we take $n$ as sufficiently large so that the chance of obtaining the Schrödinger-cat state for $P$ is negligible), so this is a deterministic hidden-variable theory. Since constancy of $\rho$ on $\Lambda_\psi^n$ is the *correct* probability distribution, according to $\lambda$-MANY, the probability of $P$, averaging over many trials, in $\lambda$-ONE, agrees with its probability in $\lambda$-MANY, in each trial, and hence with the Born rule. Therefore, it must violate Bell inequalities for certain directions of parameter settings.[17]

It seems that $\lambda$-ONE, like pilot-wave theory, must fail parameter independence in the sense of (7). And surely, parameter independence conditional on $\lambda$ clearly does *not* hold – or rather, it simply fails to apply. The $\lambda$ on the two sides of Eq.(7) cannot be identical. On the LHS, it is a microstate from an expansion adapted to $P_s^a \otimes P_t^b$; on the RHS, it is a microstate from an expansion adapted to $P_s^a \otimes P_t^{b\prime}$. Since $P_t^b$ and $P_t^{b\prime}$ differ macroscopically, they are disjoint and have no eigenstates in common (here and in the sequel, we ignore the Schrödinger-cat state).

To spell this out, the two sets of spin microstates for Bob's two choices of parameter settings, are of the form

$$\Lambda_\psi^{n,a,b} = \left\{ \phi_s^{a,k} \otimes \chi_t^{b,k};\ k = 1, \dots, n;\ \ s, t = +1, -1 \right\} \tag{8}$$

$$\Lambda_\psi^{n,a,b\prime} = \{ \phi_s^{a,k} \otimes \chi_t^{b',k};\ k = 1, \dots, n;\ \ s, t = +1, -1 \} \tag{8'}$$

[17] However, the uniformity of $\rho$ in $\lambda$-ONE cannot be justified by INVARIANCE: the proofs of Short (2023) and Saunders (2025) break down if just one among the states in a superposition is distinguished, as in $\lambda$-ONE.

where $P_s^a \phi_s^{a,k} = \phi_s^{a,k}$, $P_{+1}^a \phi_{-1}^{a,k} = 0$, etc. They have no microstates in common. $\lambda$-ONE is *ontologically* contextual;[18] the hidden-variable $\lambda$ itself depends on the context, and if Bob changes his parameter settings, he must change $\lambda$ as well.

May Bob change $\lambda$ in this way, consistent with $LOC$? Observe that these microstates amount to a pair of hidden-variables, of the form $\phi_s^a$ for Alice and $\chi_t^b$ for Bob, in a fully separable description, and that there is a 1:1 correspondence between the microstates for Alice's outcomes, in the two ensembles; in particular, the fraction of microstates with $s = +1$ is the same, in (8) and (8'), as likewise with $s = -1$ (we know this is so from $\lambda$-MANY). It follows that Bob, in mysteriously switching between (8) and (8') (which we suppose occurs with his choice of instrument setting), need not alter Alice's probabilities, nor need it change her microstate in region 3 of Fig.1. It follows that $\lambda$-ONE is consistent with $LOC$ after all, despite being a deterministic hidden-variable theory, replicating quantum probabilities. And, we belatedly see, it further satisfies factorizability, Eq.(6).

It is too good to be true. The difficulty, of course, is that $\lambda$-ONE violates measurement independence. The hidden variable $\lambda$ produced at the source depends on Alice and Bob's instruments settings, as does the measure $\rho, \Lambda$ in Eq.(5). Let us put this in terms of retrocausation: Alice and Bob's free choices retrocausally dictate the kind of microstate produced, $\rho$, and $\Lambda$, but in such a way, as just indicated, to involve no violation of $LOC$, even though violating Bell inequalities.

Our serious theory of probability is $\lambda$-MANY. From that perspective, there is good news from $\lambda$-ONE: there is no underlying action at a distance, in the sense of $LOC$, at the level of individual microstates, so none such, on taking the superposition, is concealed. But is there not bad news, too? – as concealed, it seems, is retrocausality. That might also hint at conspiracy.

I suggest not. In $\lambda$-MANY, the microstates used in defining the probabilities of Alice and Bob's outcomes, appropriate to their choice of settings, do not have to be retrodicted to the source, unlike in $\lambda$-ONE. There is no reason to do so.[19] The causal story is the usual one: $\psi$ unitarily evolves from the source to the moment of Bob's measurement, where, depending on his choice of instrument settings, and given the careful engineering of his apparatus, and given that he actually chooses to carry out his experiment, branching occurs, and a superposition of measurement outcomes results. This is the only dynamics that is operating (the Schrödinger equation evolving forward in time). Microstates only come into the picture for defining the probabilities of those outcomes, once the settings are chosen, not for tracing their causal provenance to the source. In $\lambda$-ONE, the two cannot be separated.

---

[18] Not just contextual in its value-assignments; as a simple corollary to Gleason's theorem, there is no non-contextual additive mapping of projectors to 0s and 1s. Bell (1966) discussed this in detail; so did Shimony (1984).

[19] It may yet be motivated as a way of introducing pre-measurement uncertainty, as in Saunders (2010, 2021); but that concerns only epistemic probability.

## 5. $\lambda$-MANY and Outcome Independence

The remaining question for $\lambda$-MANY is outcome independence, Eq,(2). Consider first the equivalent condition, completeness, Eq.(3). Suppose that $\psi$ is a product state of the form $\psi = \phi \otimes \chi$. Consider the probability of $P_s^a \otimes P_t^b$ in $\psi$. We suppose we can expand $\phi$ and $\chi$ in equiamplitude microstates that diagonalise $P_s^a$ and $P_t^b$ respectively, as before; but now we keep track of their numbers, say $n_a$ and $n_b$ respectively. Of these, let $m_a$ have eigenvalue +1 for $P_s^a$, and let $m_b$ have eigenvalue +1 for $P_t^b$, with all the rest eigenvalue 0. That is, on choosing a convenient ordering, we consider the two expansions

$$\phi = \phi_{+1}^{1} + \cdots + \phi_{+1}^{m_a} + \phi_{-1}^{m_a+1} + \cdots + \phi_{-1}^{n_a} \qquad (9)$$

$$\chi = \chi_{+1}^{1} + \cdots + \chi_{+1}^{m_b} + \chi_{-1}^{m_b+1} + \cdots + \chi_{-1}^{n_b}. \qquad (9')$$

The product $\phi \otimes \chi$ then consists of the sum of $n_a \cdot n_b$ equiamplitude microstates, each an eigenstate of $P_s^a \otimes P_t^b$, the $j \cdot k^{\text{th}}$ of which is one or other of:

$$\phi_{+1}^{j} \otimes \chi_{+1}^{k};\ \phi_{+1}^{j} \otimes \chi_{-1}^{k};\ \phi_{-1}^{j} \otimes \chi_{+1}^{k};\ \phi_{-1}^{j} \otimes \chi_{-1}^{k}. \quad (10)$$

Of these $m_a \cdot m_b$ have eigenvalue +1, the rest 0, so $P_s^a \otimes P_t^b$ has probability

$$p_{a,b}(s,t|\psi) = \frac{m_a \cdot m_b}{n_a \cdot n_b}. \qquad (11)$$

Similarly, the probabilities of $P_s^a \otimes I$ and $I \otimes P_t^b$ are

$$p_{a,b}(s|\psi) = \frac{m_a}{n_a},\ p_{a,b}(t|\psi) = \frac{m_b}{n_b}.$$

Their product is Eq.(11), so completeness, Eq.(3), is satisfied, hence too outcome independence Eq.(2): the probability of Alice's outcome spin-up, conditional on Bob's outcome spin-down, is the same as that conditional on Bob's outcome spin-up -- and is the same if Bob performs no measurement at all. But this reasoning has nothing to do with explaining correlations between Alice and Bob, no more than does outcome independence. On the contrary, there *are* no correlations, when completeness is satisfied.

This same reasoning shows when, precisely, completeness *is* violated: when $\psi$ is *not* a product state. It will still have expansions in terms of product states of the form (10), adapted to $P_s^a \otimes P_t^b$, but it will no longer be possible to write these expansions as the product of two expansions (9), (9') – which is what completeness requires. The microstates may be paired with one another in a more interesting way – for example, half with Alice spin-up and Bob spin-down, and half with Alice spin-down and Bob spin-up. This is the case when Alice and Bob measure spin in the same direction $a = b$, for the singlet state of spin.

That pairings like this should exist, all in a superposition, that cannot be written as the product of two sets of states, each in a superposition, is fundamental to quantum mechanics, and the way that correlations there arise. It is entanglement. Whether it calls for explanation presumably depends on the pairings.[20] In the EPRB setup there *is* a common cause: the entanglement is prepared by a device in the causal past of Alice and Bob, and it is precision-engineered to build in these special pairings, so there is an explanation. There is no distant action, but there is distant correlation--correlations that are preserved over large distances to be sure, but correlations by design. It is just a mistake to think that if $\psi$ explains some correlation between Alice and Bob, then completeness must be satisfied. It is different for factorizability Eq.(6) for a deterministic hidden variable theory, which explains, indeed (in Bell's words) *locally* explains, correlations – and which is true of the microstates in $\lambda$-MANY, considered separately, as in $\lambda$-ONE.

Completeness is equivalent to outcome independence; when $\psi$ is entangled, $\lambda$-MANY must violate the latter as well. Were there a unique outcome to Bob's experiment, then it would be a causal, law-like change, brought about by his free choice in conducting the experiment, no matter that he did not control the outcome. $LOC$ would then say that causal change cannot change Alice's probabilities -- this is outcome independence. But if Bob's experiment causes a superposition of outcomes to obtain, the verdict is just as clear: Alice's probabilities should be unchanged conditional on this superposition.

It is a good question as to how all this is to be made out without recourse to quantum mechanics. Granted, outcome independence no longer follows from $LOC$, if Bob's remote experiment does not have a unique outcome; but then what does follow from $LOC$ when he carries out his measurement? – how is this to be expressed in the language of elementary probability theory? It cannot be that Alice's probabilities should be unchanged conditional on $t = +1 \wedge t = -1$, for that is a logical contradiction.

---

[20] I am unpersuaded that it needs to be explained in general, and still less eliminated, as in the search for a separable formalism for quantum mechanics (especially not if it complicates or undermines a state-based ontology). Yet there is clearly something there; see Bédard (2026).

I suggest we take our cue from the analogous question for that other loophole, the violation of measurement independence. How is that expressed, in our elementary notation? The answer as we saw is that $\lambda$ must be written as a function of the instrument settings. Similarly, I suggest, we write outcomes $t = \pm 1$ as functions of their probabilities. The two outcomes do not exist simpliciter; they exist with certain probabilities. From statements like this, Bell inequalities cannot be derived. But there are other ways forward.

There is the reverse question posed for quantum mechanics. Outcome independence is no longer implied by $LOC$, but it remains perfectly well-defined in Everettian theory. Since equivalent to completeness, we know it is violated by entanglements, but can it be given a more direct meaning, in Everettian terms?

Outcome independence would hold if Alice's probability of outcome $s$ on measuring spin component $a$, conditional on one of Bob's outcomes, say $t = +1$, made no change to her probability. But that is Alice's probability for *s in Bob's spin-up world*, the fraction of microstates with outcome $s$ for Alice's measurement, of all those microstates with $t = +1$ outcome for Bob. There is no reason why that fraction should be the same as the fraction in Bob's spin-down world – determined by an entirely disjoint set of microstates – or the same as the fraction of all microstates, the unconditional probability for Alice's outcome$s$.

## 6. What Bell did

There are several lessons to be drawn from the observed violations of Bell inequalities. One concerns the *way* they are violated, as the parameters are varied; when the inequalities are saturated, the change in correlation functions as the angle $\theta$ is changed (leading to violations) is quadratic in $\theta$. If the probability space were a simplex, it would necessarily be linear in $\theta$. Bell pointed this out (1987 p.81-86); so did Redhead (1987), Pitowsky (1989). The Block sphere, in quantum mechanics, is not a simplex; neither, on variation of $a$, is $\Lambda_{\psi}^{n;a}$.

But once we assume ordinary quantum mechanics, the significance of Bell-inequality violations is rather different. It is a marker of entanglement, since any product state satisfies completeness, as we saw in Section 5, and all states satisfy parameter independence. The recent interest in inequality-violating probabilities as measured in high-energy physics experiments is mainly directed to tests for entanglement. However, there are other signatures of entanglement in these regimes that may be more easily detected, in particular the Peres-Horodecki criterion; see e.g. Barr et al (2025).

Violations of Bell inequalities have also been obtained for systems and experimenters separated by large distances (across Lake Geneva, for example). That is noteworthy, but of course evidence that entanglements are preserved over vastly greater distances is independently available – for example, in interference effects of individual photons traversing opposite sides of galaxies, as in gravitational lensing. That rather dwarfs terrestrial examples.

The real importance of the evidence of violations of Bell inequalities is the generality of the arguments used to derive them. In Section 2 we concluded, but with little fanfare:

$$(LOC \wedge IND \wedge UNIQUE) \rightarrow BELL\,. \qquad (12)$$

Here $IND$ is measurement independence, the banal assumption that we are free to choose our instrument settings, independent of the state of the spin-systems, as prepared at the source. $UNIQUE$ is the equally banal assumption that remote experiments have unique outcomes. Both conditions are part of our pre-theoretic understanding of the world, independent of quantum theory. It is true that the negation of $UNIQUE$ is *not* of this form – we hardly know what it *means* for uniqueness to fail, without a more detailed theory (witness our difficulty in expressing it predicatively)– but it is not obviously self-undermining (the non-uniqueness, after all, concerns *remote* experiments), and one could say the same of the negation of $IND$. Given which, and given that the violation of Bell inequalities is a scientific fact, (12) may better be cast as the logically equivalent sentence: [21]

$$(LOC \wedge IND \wedge \neg BELL) \rightarrow \neg UNIQUE. \qquad (13)$$

Were Everettian quantum theory itself ruled out by $LOC$, the inference would be, at best, suggestive ('search for a new theory in which remote outcomes are not unique, consistent with $LOC$ and $IND$'). But since, as we have seen, $\lambda$-MANY is consistent with $LOC$ and $IND$, the inference is a new and very general argument for Everett, at least as based on $\lambda$-MANY.

We should consider the alternatives. (13) is also logically equivalent to

$$(LOC \wedge UNIQUE \wedge \neg BELL) \rightarrow \neg IND. \qquad (14)$$

[21] Lev Vaidman has long argued that from quantum experiments, no action at a distance, and no retrocausation, many-worlds follows; see Vaidman (2026). However, his argument (here and elsewhere with similar effect) relies on the (deterministic) GHZ experiment, not the EPRB experiment, and says nothing about physical probabilities other than 0 and 1 (indeed there are no such things, according to Vaidman). Waegell and McQueen (2020) make a similar argument, also based on the GHZ state, rather than Bell; for a reply, see Faglia (2024).

Considered as an argument, it suffers from the difficulty that we lack the analogue of Everettian quantum theory in explaining the consequent: what does it mean, to live in a world in which $\neg IND$ is true, for example, a world in which our present choices change the past? There are suggestive examples, to be sure, but like $\lambda$-ONE, they are highly stylised. General principles remain to be found. (14) suggests a research programme, whereas (13) is an argument for a well worked-out theory.

Granted $\neg BELL$, the only remaining argument of interest is:

$$(UNIQUE \wedge \neg BELL \wedge IND) \rightarrow \neg LOC.$$

It is on better ground than (14), in that there is at least one worked-out one-world quantum theory that involves action at a distance, namely, pilot-wave theory. But it is not that well worked-out, as it is still largely restricted to non-relativistic quantum mechanics, unable, for example, to accommodate photons. Moreover, with action at a distance, $\neg LOC$, comes a preferred frame, compromising relativity theory. Given which, it is hard to see how rejection of LOC connects with Bell's concerns, or Einstein's, once $\neg UNIQUE$ is clearly on the table. Yet I cannot help but feel that this is the answer that Bell would have liked the least; Einstein, I'm not so sure.

## Acknowledgments

My thanks to Chares Bédard, Dorje Brody, Harvey Brown, Jeremy Butterfield, Wayne Myrvold, James Read, Cristi Stoica, Paul Tappenden, Christopher Timpson, and David Wallace, for helpful discussions. Errors that remain are entirely my own. I am also grateful to Alyssa Ney, without whom this piece would not have been written.